\documentclass[sigconf]{acmart}
\AtBeginDocument{%
  }

\copyrightyear{2026}
\acmYear{2026}
\setcopyright{cc}
\setcctype{by}
\acmConference[MM '26] {Proceedings of the 34th ACM International Conference on Multimedia}{November 10--14, 2026}{Rio de Janeiro, Brazil.}
\acmBooktitle{Proceedings of the 34th ACM International Conference on Multimedia (MM '26), November 10--14, 2026, Rio de Janeiro, Brazil}
\acmISBN{979-8-4007-2213-4/2026/11}
\acmDOI{10.1145/3767308.3835705}

\usepackage{soul}
\usepackage{multirow}
\usepackage{enumitem}
\usepackage{balance}

\begin{document}

\title{3DRAG: 3D-Native Object Drag Editing with Generative Models}

\author{Kuixiang Shao}
\authornote{Both authors contributed equally to this research.}
\orcid{0009-0004-9605-7892}
\email{shaokx2025@shanghaitech.edu.cn}
\affiliation{%
  \institution{ShanghaiTech University}
  \city{Shanghai}
  \country{China}
}

\author{Yuyang Jiao}
\authornotemark[1]
\orcid{0009-0000-5767-5808}
\affiliation{%
  \institution{ShanghaiTech University}
  \city{Shanghai}
  \country{China}
}
\email{jiaoyy2022@shanghaitech.edu.cn}

\author{Chenghao Zhang}
\orcid{0009-0007-9789-3341}
\affiliation{%
  \institution{ShanghaiTech University}
  \city{Shanghai}
  \country{China}
}
\email{zhangchh2023@shanghaitech.edu.cn}

\author{Wenbo Zhao}
\orcid{0009-0009-7408-0414}
\affiliation{%
  \institution{ShanghaiTech University}
  \city{Shanghai}
  \country{China}
}
\email{zhaowb2024@shanghaitech.edu.cn}

\author{Yinuo Bai}
\orcid{0009-0000-3434-212X}
\affiliation{%
  \institution{ShanghaiTech University}
  \city{Shanghai}
  \country{China}
}
\email{baiyn2022@shanghaitech.edu.cn}

\author{Jingyi Yu}
\authornote{Corresponding Author}
\orcid{0000-0002-8580-0036}
\affiliation{%
  \institution{ShanghaiTech University}
  \city{Shanghai}
  \country{China}
}
\email{yujingyi@shanghaitech.edu.cn}

\renewcommand{\shortauthors}{Kuixiang Shao et al.}

\begin{abstract}
The rapid progress of 3D generative models has made 3D asset creation increasingly accessible. However, achieving intuitive, precise, and interactive user control, such as point-based drag editing, remains challenging. Existing 3D drag methods either rely on slow 2D diffusion priors, such as Score Distillation Sampling, or directly warp geometry, often causing multi-view inconsistency and topological artifacts. In this paper, we present 3DRAG, a training-free framework that executes precise drag editing natively within the explicit 3D latent space of large-scale generative models, without SDS-based optimization or iterative point tracking. To ensure geometrically plausible deformations, 3DRAG explicitly extends the drag displacement field into 3D space, incorporating an Occupancy-Aware Winner-Takes-All (OA-WTA) strategy to robustly handle sparse geometries. Furthermore, we introduce a generative editing pipeline that explicitly transforms latent representations via the 3D displacement field. By fully leveraging the model's generative capabilities, it achieves smooth geometric deformations while maintaining the object's identity. To further transfer original geometry and texture with high fidelity, we propose Logits-Level Token Fusion, which accurately injects source information into the destination region. To evaluate 3D point-based editing, we construct a novel dataset featuring diverse 3D assets and complex multi-point drag instructions. Quantitative and qualitative experiments demonstrate that 3DRAG achieves state-of-the-art performance, surpassing existing baselines in drag accuracy, geometric fidelity, and unedited region preservation.\footnotemark

\footnotetext{Our code is publicly available at \url{https://github.com/L-avenir/3DRAG}.}
\end{abstract}



\begin{teaserfigure}
  \vspace{-2mm}
  \includegraphics[width=\textwidth]{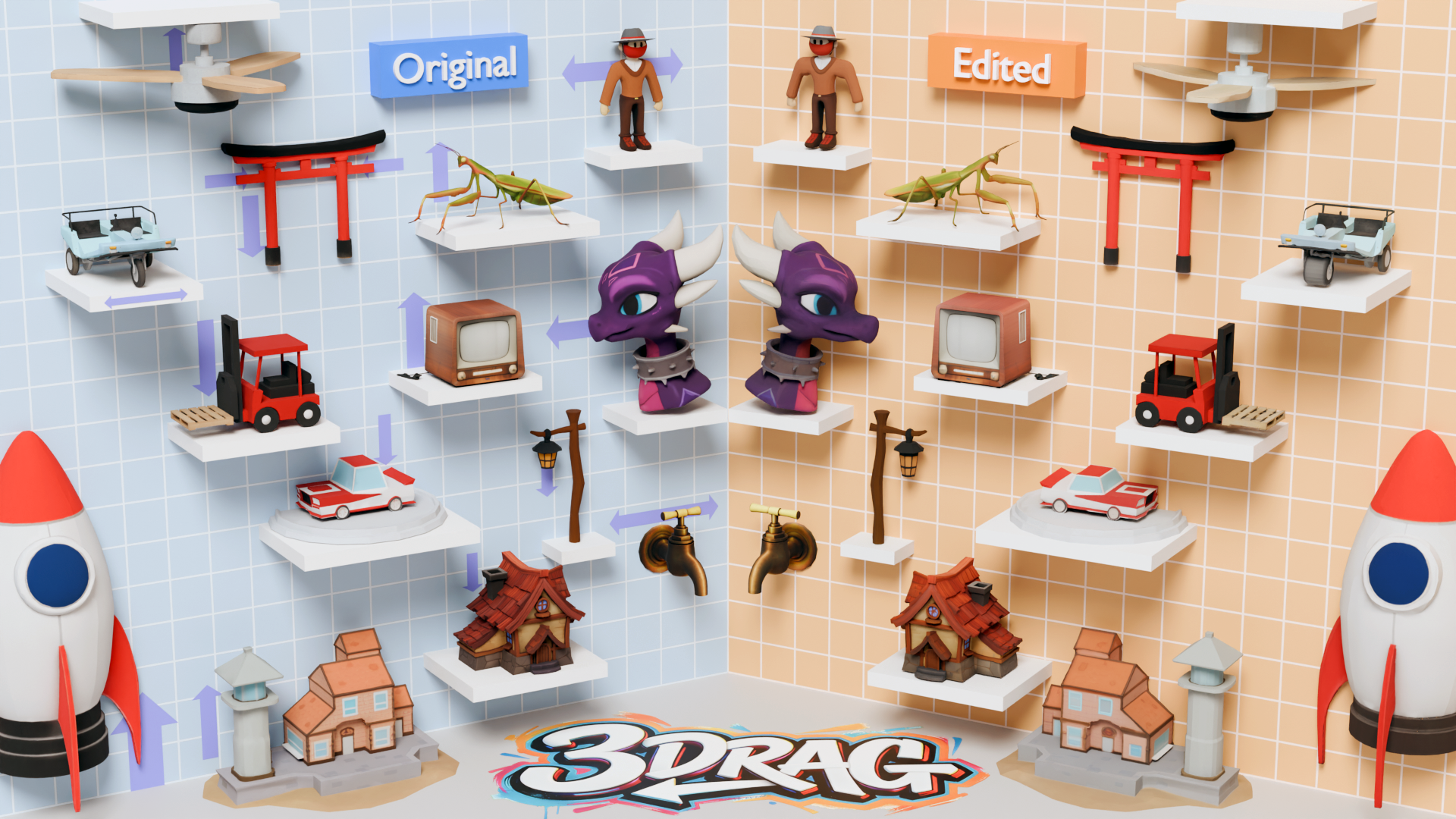}
  \vspace{-8mm}
  \caption{We propose 3DRAG, a novel framework that achieves precise, high-fidelity 3D-native drag editing without SDS or per-edit optimization. It enables users to intuitively perform natural, view-consistent spatial manipulations while preserving the identity of the unedited region.}
  \label{fig:teaser}
\end{teaserfigure}

\maketitle

\section{Introduction}

The recent advancements in 3D generative models~\cite{gao2022get3d, xiang2025structured, zhang2024clay,li2025triposg} have democratized 3D content creation, demonstrating impressive capabilities in synthesizing high-fidelity 3D assets from text or image prompts. However, for downstream applications like game development, virtual reality, and robotics, generation alone is insufficient; users frequently demand fine-grained, highly controllable manipulation over the generated content. While existing semantic editing methods~\cite{livoxhammer,chi2026vinedresser3d, li2025cmd} excel at localized replacement and stylization, they inherently struggle to execute precise geometric deformations. To bridge this gap, drag-based editing provides an interaction paradigm similar to conventional 3D modeling workflows, empowering users with intuitive, point-driven control for precise spatial manipulation.
    
Inspired by drag-based 2D image editing techniques~\cite{pan2023drag,shi2024dragdiffusion,yin2025lazydrag,zhao2024fastdrag,koo2025flowdrag,zhang2025gooddrag}, researchers have extended this paradigm into the 3D domain. Existing 3D drag frameworks generally follow two paradigms: optimizing 3D representations\cite{qu2025drag} (e.g., meshes or 3DGS~\cite{kerbl3Dgaussians}) via Score Distillation Sampling (SDS)\cite{pooledreamfusion}, or explicitly warping 3D geometry followed by a 2D diffusion refinement process~\cite{dong3dgs,yoo2024apap}. Despite their potential, both pipelines share a critical bottleneck: a heavy reliance on 2D diffusion priors~\cite{salimans2022progressive,ruiz2023dreambooth}. Inherently lacking true 3D awareness, these approaches suffer from computationally expensive iterative optimization and are highly prone to multi-view inconsistencies (e.g., the Janus problem~\cite{hong2023debiasing}). Furthermore, forcing 2D priors to guide 3D structural changes frequently leads to severe geometric violations, including unnatural stretching and topological artifacts.

Our key insight is that to achieve highly consistent and plausible deformations, the manipulation should be executed directly within the native 3D space, bypassing the reliance on 2D priors. Driven by this insight, we propose 3DRAG, a novel training-free framework that achieves precise, high-fidelity 3D drag editing without SDS-based optimization or iterative point tracking. Our pipeline seamlessly integrates spatial transformation with the native generative process. First, we construct a 3D Displacement Field by extending an elasticity-based model into the 3D domain. Augmented with an Occupancy-Aware Winner-Takes-All (OA-WTA) strategy, this formulation robustly handles the sparse feature structures of modern 3D generation models and naturally circumvents the topological tearing and geometric inconsistencies that frequently plague optimization-based approaches. Second, we execute an identity-preserving generative editing process. The input object is inverted to its latent representation and spatially warped according to the displacement field. During the subsequent denoising steps, we introduce a synergistic mechanism of QKV replacement and Logits-Level Token Fusion. This targeted intervention smoothly transfers texture and geometric details to the new 3D coordinates, guaranteeing seamless blending in the newly occupied regions while preserving the unedited region. Consequently, 3DRAG enables users to intuitively perform complex, view-consistent spatial manipulations without sacrificing the original object's identity.

To rigorously evaluate the effectiveness of our method, we construct a novel and comprehensive dataset featuring diverse 3D assets paired with complex multi-point drag instructions. Quantitative and qualitative evaluations against representative mesh-based and 3DGS-based baselines demonstrate the clear superiority of our approach. 3DRAG achieves state-of-the-art performance across drag accuracy, unedited region preservation, and overall 3D quality. Our main contributions are summarized as follows:

\begin{itemize}[leftmargin=*]
\item We propose 3DRAG, a novel training-free framework for 3D-native object drag editing that bypasses unstable 2D prior optimization.
\item We formulate a 3D displacement field equipped with an Occupancy-Aware Winner-Takes-All (OA-WTA) strategy to handle the inherent sparsity of modern 3D latents.
\item We introduce an identity-preserving generative pipeline featuring a novel Logits-Level Token Fusion mechanism to effectively decouple coordinate-feature entanglement.
\item We construct a new dataset for 3D drag tasks, where 3DRAG significantly outperforms existing mesh-based and 3DGS-based baselines.
\end{itemize}

\section{Related Works}

\subsection{3D generative models}

Early 3D generation methods~\cite{metzer2023latent,chen2023fantasia3d,wang2023prolificdreamer}, such as DreamFusion~\cite{pooledreamfusion} and Magic3D~\cite{lin2023magic3d}, were largely built upon 2D diffusion priors and Score Distillation Sampling (SDS), which optimize implicit 3D representations such as NeRFs~\cite{mildenhall2021nerf} and, more recently, 3D Gaussian Splatting~\cite{kerbl3Dgaussians}. Later analyses and variants~\cite{liang2024luciddreamer,qiu2024richdreamer} attempted to stabilize SDS while mitigating over-smoothing and Janus-like multi-face artifacts. Multi-view diffusion models~\cite{liu2023zero,long2024wonder3d,wang2023imagedream,lu2024direct2}, exemplified by MVDream~\cite{shimvdream}, further improve multi-view consistency by learning view-correlated priors or reconstructing 3D from multi-view predictions.

\begin{figure*}[t]
    \centering
    \vspace{-2mm}
    \includegraphics[width=\textwidth]{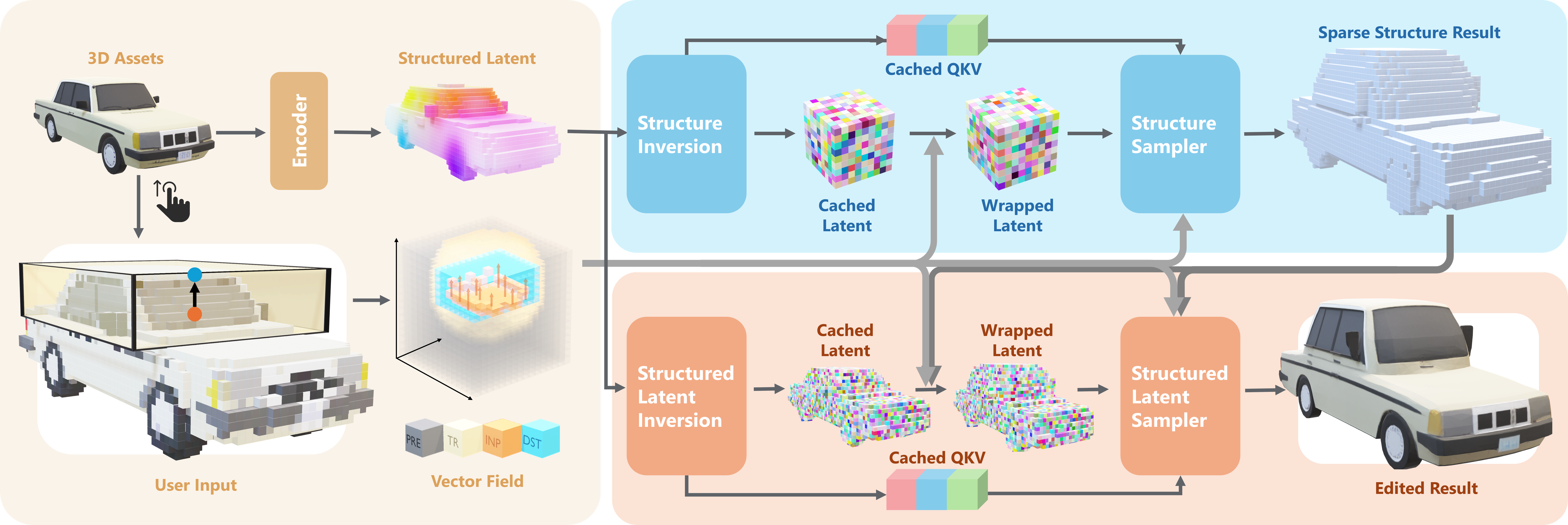}
    \vspace{-8mm}
    \caption{3DRAG pipeline. Given an input object, our method first encodes it into a sparse voxel latent while simultaneously calculating a displacement field based on user drag instructions. Following the TRELLIS architecture, the core editing process begins with an inversion in both the Structure (ST) and Structured Latent (SLAT) stages to acquire the cached latent $\tilde{z}_T$ and attention features $\tilde{Q}, \tilde{K}, \tilde{V}$. The sampling process is guided by the cached representations and the displacement field, enabling natural geometric deformation while preserving the object’s original details and unedited regions.}
    \label{fig:method}
    \vspace{-4mm}
\end{figure*}

More recently, native 3D and feed-forward generative models have emerged as a more efficient alternative by directly learning 3D representations without test-time optimization. Early examples include GET3D~\cite{gao2022get3d}, which generates textured meshes, and large reconstruction models~\cite{honglrm,zhang2024gs,xu2024instantmesh,TripoSR2024} which predict 3D structures from one or a few images in a single forward pass. Building on this trend, CLAY~\cite{zhang2024clay} learns a controllable 3D-native geometry-and-material generator with a multi-resolution VAE and latent DiT~\cite{peebles2023scalable}; TRELLIS~\cite{xiang2025structured} introduces Structured Latents (SLAT) that can be decoded into radiance fields, 3D Gaussians, or meshes; and TripoSG~\cite{li2025triposg} further scales rectified-flow-based 3D shape synthesis to high-fidelity mesh generation. Although these approaches substantially improve efficiency and output quality, they lack the interactive mechanisms necessary for precise, fine-grained geometric manipulation.

\subsection{Drag-based Editing}

Drag-based image editing aims to provide intuitive and precise spatial control by directly manipulating user-specified points. Early works formulate this problem on the latent manifold of GANs, where point correspondences are enforced via iterative feature supervision and latent optimization~\cite{pan2023drag}. This paradigm has been extended to diffusion models by incorporating feature-level guidance into the denoising process, enabling more generalizable editing on real images~\cite{shi2024dragdiffusion,moudragondiffusion,mou2024diffeditor}. To improve robustness and usability, subsequent approaches focus on stabilizing point tracking, reducing optimization ambiguity, and enhancing semantic consistency through improved feature matching, confidence modeling, or region-aware constraints~\cite{ling2024freedrag,cui2024stabledrag,liu2024drag}. Meanwhile, a parallel line of work seeks to accelerate editing inference by replacing iterative optimization with approximate or one-step solutions, such as latent warping, optimization-free generation, or learned motion priors~\cite{zhao2024fastdrag,shin2024instantdrag,koo2025flowdrag,shi2025lightningdrag}. Despite these advances, existing methods are inherently designed for 2D image manipulation; naively applying them to multi-view settings often leads to inconsistent geometry and view-dependent artifacts.

\subsection{3D Editing}

3D editing has evolved from NeRF-based local manipulation to explicit 3DGS editing and, more recently, native 3D latent-space editing. Early NeRF-based methods~\cite{liu2021editing, mirzaei2023spin,yin2023ornerf,zhuang2024tip,park2024ed,chen2023neuraleditor,haque2023instruct} typically propagate user instructions such as text, images, masks, or scribbles from rendered views back to the underlying radiance field, enabling object removal, re-texturing, shape modification, and local insertion while aiming to preserve cross-view consistency. With the emergence of 3D Gaussian Splatting, subsequent methods exploit its explicit representation to support faster and more localized edits, often combining multi-view diffusion priors, depth-aware guidance, and consistency regularization to improve visual quality and reduce view conflicts~\cite{chen2024gaussianeditor, wu2024gaussctrl, chen2024dge, wen2025intergsedit, sun2024gseditpro}. In parallel, drag-based editing extends point-based interaction to 3D geometry by allowing users to specify handle-target constraints for deformation. Representative approaches deform meshes or Gaussians explicitly and then rely on diffusion- or reconstruction-based priors to complete the edited content. However, they remain challenged by large deformations and structural preservation~\cite{yoo2024apap, qu2025drag, dong3dgs}. More recently, native 3D latent-space editing has emerged as an alternative to 2D prior distillation, directly manipulating the latent variables of 3D generative models to improve edit coherence and preserve unedited region~\cite{livoxhammer, chi2026vinedresser3d, li2025cmd}.

\section{Method}
\label{sec:method}

We present 3DRAG, a novel and efficient framework for precise, geometrically plausible 3D drag-based editing by executing spatial manipulations natively in the explicit 3D space of modern generative models. As illustrated in Fig.~\ref{fig:method}, our pipeline decomposes editing into two stages. In Sec.~\ref{sec:preliminary}, we briefly review the foundations of native 3D generative models and the principles of elasticity-based displacement. In Sec.~\ref{sec:displacement_field} we describe the construction of the 3D drag displacement field, showing how user drag instructions (source handles, target locations, and an editable mask) are translated into a deterministic spatial deformation. In Sec.~\ref{sec:edit_process} we introduce an identity-preserving generative editing pipeline, which applies the derived displacement field to latents and attention representations, realizing geometric deformation while rigorously preserving the object’s original details.

\begin{figure}[t]
    \centering
    \vspace{-2mm}
    \includegraphics[width=0.48\textwidth]{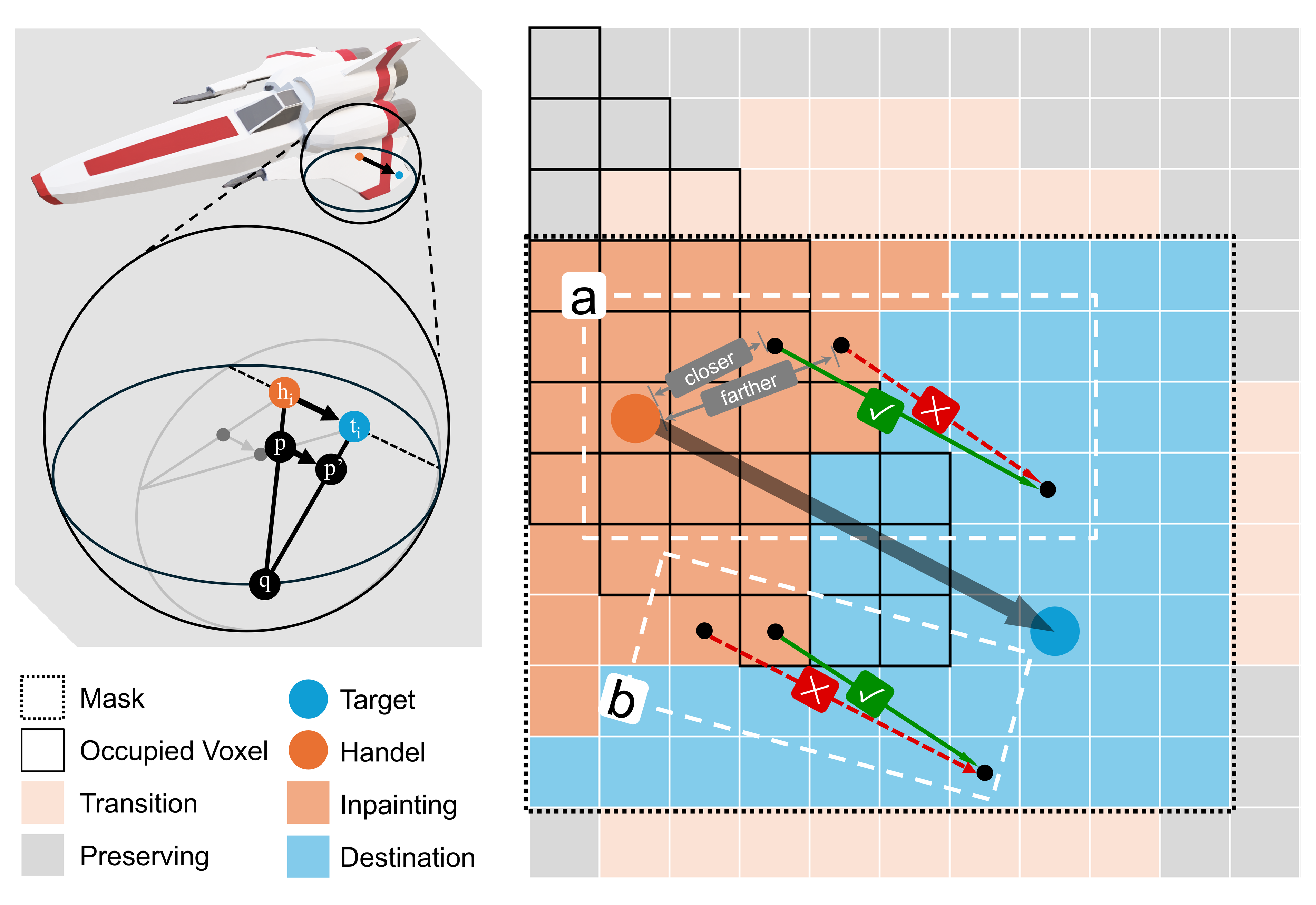}
    \vspace{-8mm}
    \caption{Construction of the 3D displacement field. (Left) Visualization of the 3D elastic deformation model. (Right) Compared to the basic WTA (box a), OA-WTA (box b) prioritizes occupied source voxels.}
    \label{fig:field}
    \vspace{-4mm}
\end{figure}

\subsection{Preliminary}
\label{sec:preliminary}
\paragraph{3D Generative Model and Inversion}

Recent advancements in 3D content creation have transitioned from optimizing 2D priors to native 3D generative models directly operating on volumetric data. In this work, we adopt TRELLIS~\cite{xiang2025structured} as our base model, which leverages the Structured Latents (SLAT) representation. By anchoring dense visual features to a sparse, active 3D voxel grid—typically operating at a resolution of $64^3$—SLAT explicitly captures complex spatial topologies. Formally, a 3D asset is parameterized as $\{(\mathbf{p}_i, \mathbf{f}_i)\}_{i=1}^N$, where $\mathbf{p}_i \in \mathbb{R}^3$ denotes the discrete coordinates of the active voxels and $\mathbf{f}_i \in \mathbb{R}^d$ represents their latent features. The generation naturally proceeds in two sequential stages: a Structure (ST) stage that establishes the sparse voxel topology, followed by a Structured Latent (SLAT) stage that populates the fine-grained features. This native 3D formulation provides an explicit canvas for precise geometric manipulations directly on $\mathbf{p}_i$ and $\mathbf{f}_i$.

To enable semantics-preserving drag editing on existing 3D assets, the input must be deterministically mapped back into the generative model's noise space. Because standard first-order ODE solvers suffer from numerical truncation errors that degrade reconstruction fidelity, we utilize an exact inversion trajectory to reverse the flow. Specifically, we integrate the rectified-flow ODE from data ($t=0$) to noise ($t=1$) using a second-order Taylor-improved Euler scheme~\cite{wang2025taming, livoxhammer}. For a given timestep $t$ and integration step $\Delta t$, the state is inverted via:

\vspace{-2mm}
\begin{equation}
    x_{t+\Delta t} = x_t + \Delta t \, \mathbf{v}_\theta(x_t, t) + \frac{1}{2} \Delta t^2 \partial_t \mathbf{v}_\theta(x_t, t),
\end{equation}

where the time derivative of the velocity field is approximated by:

\vspace{-2mm}
\begin{equation}
    \partial_t \mathbf{v}_\theta(x_t, t) \approx \frac{\mathbf{v}_\theta(x_{t+\Delta t/2}, t + \Delta t/2) - \mathbf{v}_\theta(x_t, t)}{\Delta t/2}.
\end{equation}

Throughout the exact inversion process, we extract and cache the intermediate structured latents $\widetilde{z}$, alongside the self-attention Query, Key, and Value tokens (denoted as $\widetilde{Q}, \widetilde{K}, \widetilde{V}$) at each timestep $t$. During the subsequent editing phase, the cached tokens provide geometric features and material details, ensuring that the identity of the edited area is preserved with high quality.

\paragraph{Elastic Deformation Basics}

To overcome the computational bottlenecks and instability, recent 2D image-editing frameworks~\cite{zhao2024fastdrag,yin2025lazydrag}recast drag manipulation as a physical material-deformation problem. Rather than iteratively optimizing latent variables and tracking control points, this paradigm explicitly simulates the behavior of a stretched elastic material. Concretely, given source handle points $\mathcal{H}=\{\mathbf{h}_k\}_{k=1}^K$, their target positions $\mathcal{T}=\{\mathbf{t}_k\}_{k=1}^K$, and a user-defined editable mask $\mathcal{M}$, the method constructs a deterministic spatial displacement field $\mathcal{D}(\mathbf{x})$. This material-based formulation yields a direct, optimization-free spatial warp that exactly enforces the handle constraints $\mathbf{h}_k+\mathcal{D}(\mathbf{h}_k)=\mathbf{t}_k$ and guarantees $\mathcal{D}(\mathbf{x})=0$ for all $\mathbf{x}\notin\mathcal{M}$. While previous works are strictly confined to 2D image grids, the physical principles governing material stretching are agnostic to specific 2D pixel priors. This motivates us to elevate this paradigm into 3D volumetric space, which we formulate in the subsequent section.

\begin{figure}[t]
    \centering
    \vspace{-2mm}
    \includegraphics[width=0.45\textwidth]{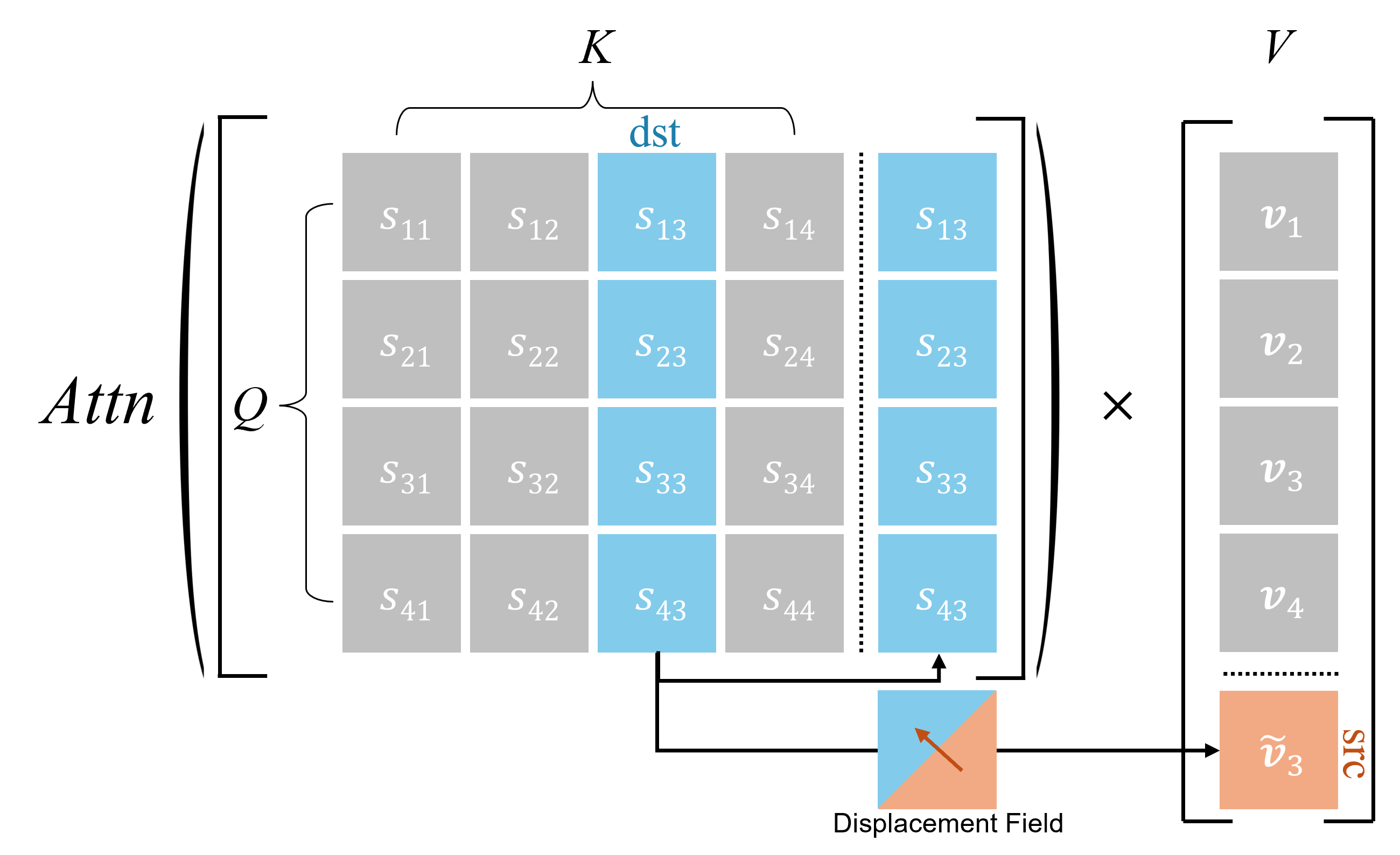}
    \vspace{-4mm}
    \caption{Logits-Level token fusion mechanism. Visualization of duplicating destination attention logits and appending the corresponding cached source values $\widetilde{V}$ based on the displacement field mapping.}
    \label{fig:logit}
    \vspace{-5mm}
\end{figure}

\begin{table*}[t]
\centering
\vspace{-2mm}
\caption{Quantitative results compared with baselines.}
\label{tab:comparison} 
\vspace{-4mm}
\setlength{\tabcolsep}{3.35mm} %
\resizebox{1.0\linewidth}{!}{
\begin{tabular}{l|cccc|cc|ccc|c}
\toprule
\multirow{3}{*}[0.8ex]{Method} & \multicolumn{4}{c|}{Unedited Region Preservation} & \multicolumn{2}{c|}{Drag Accuracy} &\multicolumn{3}{c|}{Overall 3D Quality} & \multicolumn{1}{c}{Time} \\
\cmidrule(lr){2-11} & CD $\downarrow$ & PSNR $\uparrow$ & SSIM $\uparrow$ & LPIPS $\downarrow$ & DAI $\downarrow$ & MD $\downarrow$ & FID $\downarrow$ & FVD $\downarrow$ & GPTEval3D $\uparrow$ & Runtime(s) $\downarrow$ \\
\midrule
3DGS-Drag & 0.023 & 24.31 & 0.912 & 0.214 & 0.262 & 18.25 & 156.74 & 295.22 & 874.90 & 752.70 \\
Drag Your Gaussian & 0.026 & 29.37 & 0.923 & 0.187 & 0.223 & 18.45 & 97.13 & 303.98 & 1101.83 & 1820.93 \\
ARAP & 0.188 & 14.95 & 0.644 & 0.322 & 0.413 & 77.38 & 247.90 & 964.33 & 671.27 & 3.32 \\
APAP & 0.047 & 22.52 & 0.918 & 0.079 & 0.368 & 18.01 & 63.31 & 181.29 & 1259.49 & 275.13 \\
\midrule
\textbf{Ours} & \textbf{0.008} & \textbf{31.50} & \textbf{0.974} & \textbf{0.025} & \textbf{0.187} & \textbf{15.89} & \textbf{21.99} & \textbf{42.32} & \textbf{1316.46} & 227.74 \\
\bottomrule
\end{tabular}
}
\vspace{-2mm}
\end{table*}

\subsection{Construction of the 3D Displacement Field}
\label{sec:displacement_field}

To transform user interactions into a geometrically plausible 3D manipulation, we explicitly construct a continuous 3D displacement field, which is then discretized at the voxel level. Unlike dense 2D image grids, native 3D representations (e.g., SLAT) are inherently sparse, making naive extensions of 2D drag-based editing often lead to spatial conflicts and topological artifacts. To address this, we extend elastic deformation to 3D space and introduce an occupancy-aware mapping strategy to robustly resolve voxel-level competition.

\paragraph{3D Elastic Deformation Model.} 
Given user-defined instructions $I = \{\mathcal{H}, \mathcal{T}, \mathcal{M}\}$, we first determine the spatial shift for each point $p = (x, y, z) \in \mathcal{M}$. For each user-specified point pair $\{h_i, t_i\}$, the drag operation is initialized with a base displacement vector $d_i = t_i - h_i$, as shown in Fig.\ref{fig:field}. Following the elastic deformation principle~\cite{zhao2024fastdrag}, we extract the minimum bounding sphere of the editable region $\mathcal{M}$. Taking $h_i$ as the origin, we cast a ray through $p$, which intersects the boundary of the sphere at point $q$. Then, the displacement at $p$ contributed by this pair of dragging points can be expressed as:

\vspace{-2mm}
\begin{equation}
    d_{p, i} = \frac{||p-q||_2}{||h_i-q||_2}d_i.
\end{equation}
According to this formulation, the displacement approaches $d_i$ when $p$ is close to the handle point $h_i$, and smoothly decays to zero as it approaches the deformation boundary.

\paragraph{Occupancy-Aware Winner-Takes-All (OA-WTA) Strategy.}
For multi-point dragging, we first determine the governing handle $h_k$ for a given source point $p_{\text{src}}$ using a spatial distance-based weight (Fig.~\ref{fig:field}, dashed box a):

\vspace{-2mm}
\begin{equation}
    w_{i} =
    \begin{cases}
    \frac{1}{\|p_{\text{src}} - h_i\|_2}, & p_{\text{src}} \neq h_i, \\
    +\infty, & p_{\text{src}} = h_i.
    \end{cases}
\end{equation}
Here, each point $p_{\text{src}}$ is assigned to the handle $h_k$ with the maximum weight.
Since our method operates on discrete voxels, projecting points onto the grid inevitably maps multiple $p_{\text{src}}$ to the same destination $p_{\text{dest}}$, leading to spatial collisions. Given the sparsity of 3D representation, relying purely on basic WTA can lead to empty voxels inappropriately overriding occupied ones, causing cavities or topological tearing on the final mesh. To address this issue, we enhance the strategy with occupancy awareness (OA-WTA), to determine the "winner" source point for each destination voxel:
\begin{equation}
    \hat{w}(p_{\text{src}}) = O(p_{\text{src}}) + \frac{1}{\|p_{\text{src}} - h_k\|_2 + 1},
\end{equation}
where $O(\cdot)\in \{0,1\}$ is a binary occupancy indicator. This formulation mathematically guarantees that occupied source points ($\hat{w}>1$) strictly dominate unoccupied ones ($\hat{w}\le 1$), ensuring that occupied source points are prioritized (Fig.~\ref{fig:field}, dashed box b). This yields a more reliable displacement field, resulting in meshes with improved topological consistency and reduced artifacts.

\paragraph{Explicit Region Partitioning.}

To condition the subsequent generative process, we leverage the displacement field to partition the 3D space into four distinct functional regions:
\textbf{(1) Destination Region ($\Omega_{\text{dest}}$):} Target areas accommodating the displaced source content. \textbf{(2) Inpainting Region ($\Omega_{\text{inp}}$):} Voids that require generative completion due to a lack of source mapping. \textbf{(3) Preserving Region ($\Omega_{\text{pre}}$):} Unedited spaces where original geometry and textures are preserved. \textbf{(4) Transition Region ($\Omega_{\text{trans}}$):} Boundary interfaces that ensure seamless blending.
These regions, together with the displacement field, provide structured guidance for controllable 3D drag-based editing.

\subsection{Identity-Preserving Generative Editing}
\label{sec:edit_process}

With the explicit 3D region partitioning and point-wise displacement field established, we introduce a generative editing pipeline. Our goal is twofold: to induce natural, artifact-free geometric deformations, and to faithfully propagate source visual features to the destination region while strictly preserving the unedited areas.

\paragraph{Latent Initialization}
To anchor the editing process in the generative space, we project the input 3D asset into the sparse Structured Latent (SLAT) space and execute the exact ODE diffusion inversion. During inversion, we cache the inverted latent $\widetilde{z}_T$ at the final timestep $T$, along with the intermediate attention Query, Key, and Value tokens ($\widetilde{Q}, \widetilde{K}, \widetilde{V}$) across all layers and timesteps.

To initialize the latent $z_T$, we perform a region-aware spatial replacement. Guided by the displacement field, the inverted source latents $\widetilde{z}_T$ are warped to the target coordinates, while the preserving region $\Omega_{\text{pre}}$ remains unchanged. The inpainting region $\Omega_{\text{inp}}$ is filled with Gaussian noise to avoid interpolation artifacts and enable native generation. Formally,  let $\mathbf{W}(p)$ denote the reverse mapping function that retrieves the source coordinate for a given spatial location $p$. The initial latent state $z_T(p)$ is defined as:
\begin{equation}
    z_T(\mathbf{p}) = 
    \begin{cases} 
        \widetilde{z}_T(\mathcal{W}(\mathbf{p})), & \mathbf{p} \in \Omega_{\text{dest}} \\
        \widetilde{z}_T(\mathbf{p}), & \mathbf{p} \in \Omega_{\text{pre}} \cup \Omega_{\text{trans}} \\
        \boldsymbol{\epsilon} \sim \mathcal{N}(\mathbf{0}, \mathbf{I}), & \mathbf{p} \in \Omega_{\text{inp}}
    \end{cases},
\end{equation}
In the Structure (ST) stage, this reassignment is applied directly. However, in the SLAT stage, coarse geometry may yield invalid correspondences. To prevent artifacts, a destination point $p\in\Omega_{\text{dest}}$ is retained only if it exists in the generated voxels of ST stage and maps to an occupied source voxel($\mathcal{W}(p)$ is not empty). Otherwise, it is reassigned to $\Omega_{\text{inp}}$.

\paragraph{Unedited Region Preservation}
The wide receptive field of global self-attention often causes unintended alterations in unedited regions. To strictly preserve the identity of the Unedited Region $\Omega_{\text{pre}}$, we explicitly intervene within the attention mechanism during the denoising steps. Inspired by prior works~\cite{cao2023masactrl,yin2025lazydrag,yin2026trainingfree}, we directly replace the Query ($Q_{\mathbf{p}}$), Key ($K_{\mathbf{p}}$), and Value ($V_{\mathbf{p}}$) tokens for all spatial coordinates $\mathbf{p} \in \Omega_{\text{pre}}$ with their cached inversion counterparts $\{\widetilde{Q}_{\mathbf{p}}, \widetilde{K}_{\mathbf{p}}, \widetilde{V}_{\mathbf{p}}\}$.

\begin{figure*}[t]
    \centering
    \vspace{-2mm}
    \includegraphics[width=\textwidth]{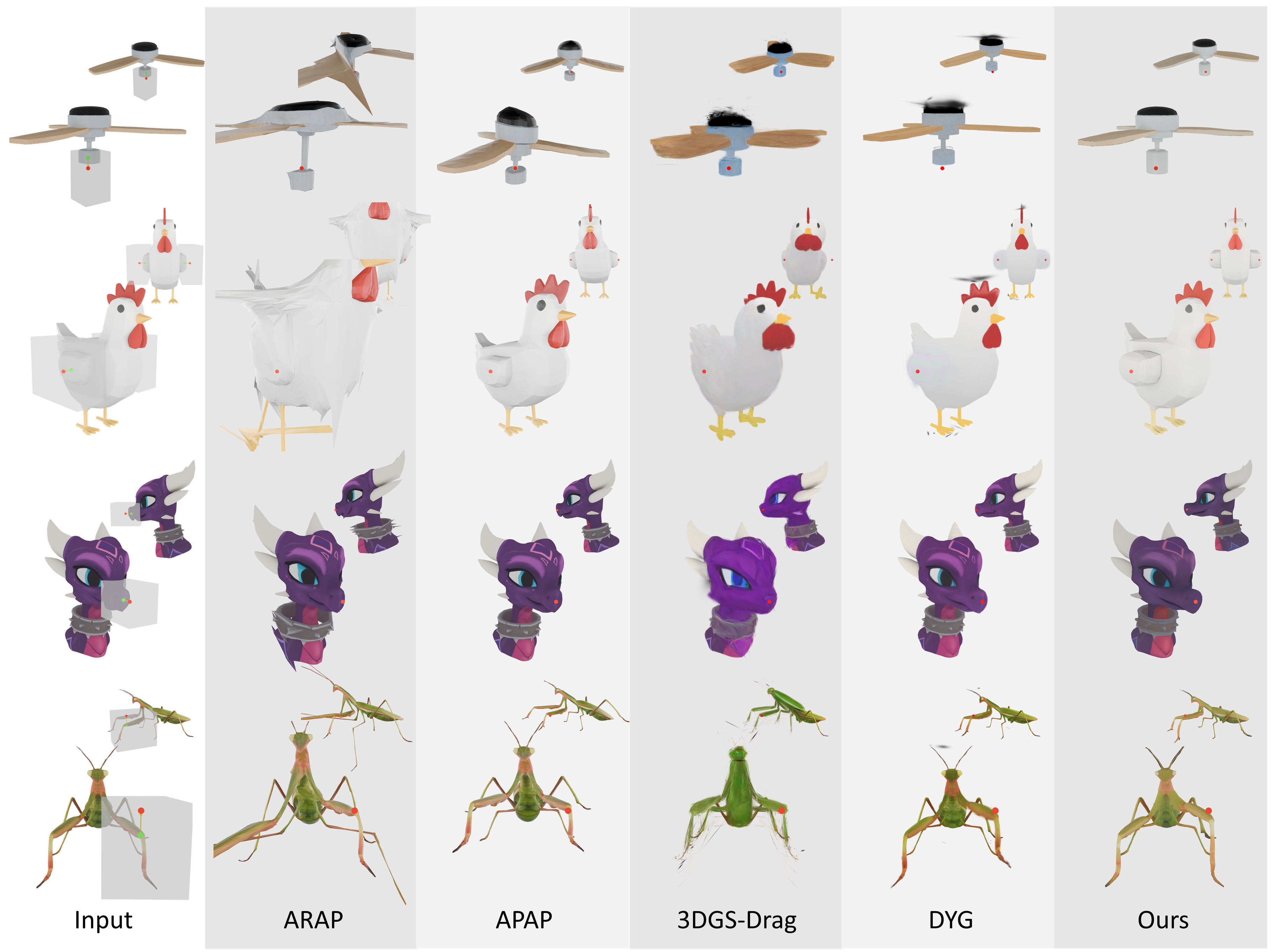}
    \vspace{-8mm}
    \caption{Qualitative results compared with baselines. The first column shows the input, highlighting the editable mask (grey), handle points (green), target locations (red), and drag directions (arrows). The subsequent columns display the results of four baselines and our method.}
    \label{fig:compare}
    \vspace{-4mm}
\end{figure*}

\paragraph{Logits-Level Token Fusion}
Generating high-fidelity textures in the destination region ($\Omega_{\text{dest}}$) requires transferring fine-grained visual features from the source structures. Conventional methods first apply Rotary Positional Embeddings (RoPE)~\cite{su2024roformer} to align cached source Keys ($\widetilde{K}_{\text{src}}$) with the destination, and then concatenate them with the current sequence ($K_{\text{new}} = [K, \widetilde{K}_{\text{src}}^{RoPE}]$), allowing destination tokens to aggregate features from the displaced source. However, modern 3D frameworks such as TRELLIS inject absolute 3D positional encodings directly into latent tokens before attention, entangling coordinates with features. As a result, destination tokens assign negligible attention to source keys under basic attention, leading to the loss of source features.

To address this, we propose \textbf{Logits-Level Token Fusion} (Figure~\ref{fig:logit}), which intervenes directly in the pre-softmax attention logits. Guided by the displacement field, we retrieve cached source Values ($\widetilde{V}_{\text{src}}$) and append them to the current Value matrix $V$. Crucially, instead of computing new attention scores for these appended tokens, we duplicate the corresponding columns of the destination tokens from the original logits matrix $L=QK^T/\sqrt{d}$. This explicitly assigns the displaced source features identical attention weights to their destination counterparts, guaranteeing that the appended source Values are proportionally integrated into the attention output regardless of their original 3D coordinates.

\section{Experiments}

\subsection{Setup}

\paragraph{Implementation Details.}

Our method is built on TRELLIS~\cite{xiang2025structured} and Voxhammer\cite{livoxhammer} and executed on a single NVIDIA H20 GPU. Unless otherwise noted, all baseline methods are evaluated using their official implementations and default hyperparameters. We operate natively within the sparse SLAT space at a resolution of $64^3$. We perform inversion using RF-Solver~\cite{wang2025taming} along the exact inversion trajectory. We set the sampling steps to 25 for both the inversion and denoising phases, and set the text classifier-free guidance (CFG)~\cite{ho2021classifierfree} scale to 5.0 for both stages to balance reconstruction fidelity and edit creativity. During the generative editing phase, our proposed Logits-Level Token Fusion and Unedited Region Preservation strategies are activated across all self-attention layers. 

\begin{figure*}[t]
    \centering
    \vspace{-2mm}
    \includegraphics[width=\textwidth]{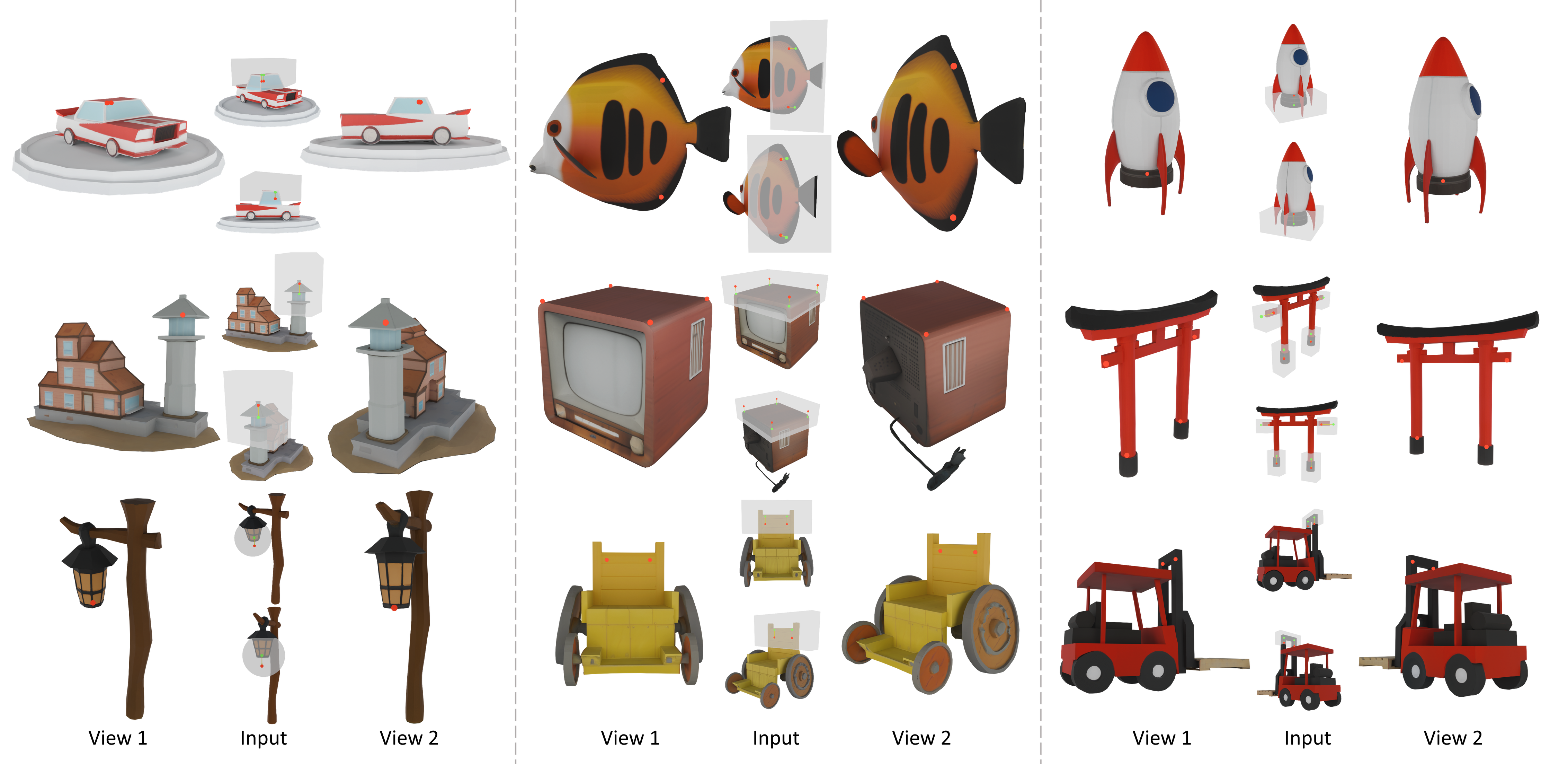}
    \vspace{-8mm}
    \caption{Gallery of 3DRAG results. For each case, the two vertically stacked images in the middle show input views with the editable mask (grey), source handles (green), and target locations (red). The images on the left and right display the corresponding edited results, with red points marking the achieved targets.}
    \label{fig:gallery}
    \vspace{-2mm}
\end{figure*}

\begin{figure}[t]
    \centering
    \includegraphics[width=0.45\textwidth]{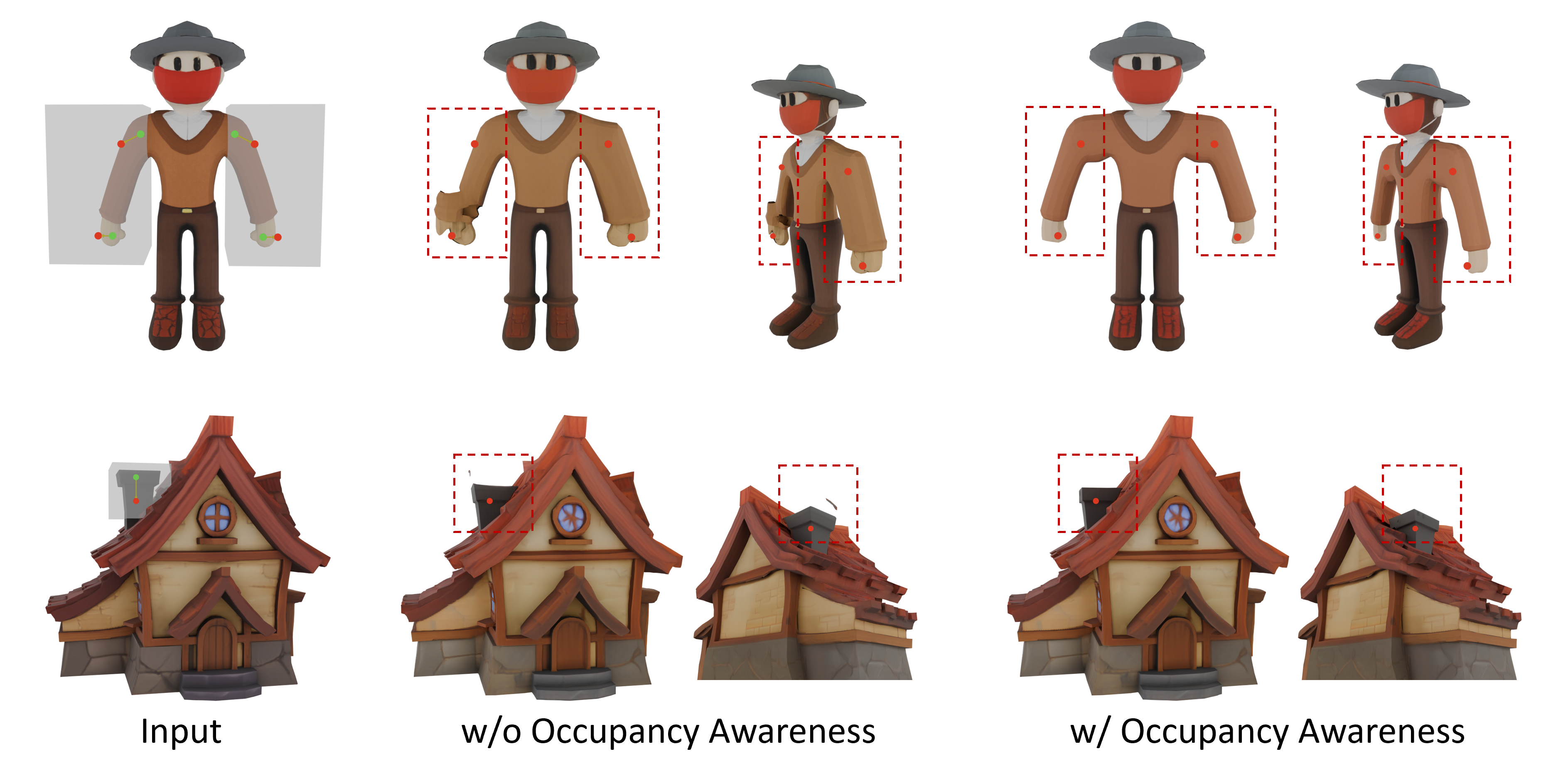}
    \vspace{-4mm}
    \caption{Visualization of OA-WTA strategy ablation. Our method effectively reduces deformation artifacts and floating fragments.}
    \label{fig:ablation_geo}
    \vspace{-5mm}
\end{figure}

\paragraph{Baselines.}

We compare 3DRAG against representative drag-based editing methods, which can be categorized into 3D Gaussian Splatting-based~\cite{kerbl3Dgaussians} and mesh-based approaches. For the Gaussian-based baselines, we evaluate 3DGS-Drag~\cite{dong3dgs} and Drag Your Gaussian~\cite{qu2025drag}. We strictly follow their official implementations and default hyperparameter settings. To ensure a fair comparison regarding spatial constraints, we identify Gaussians within the user-defined mask as editable. For the mesh-based baselines, we compare against the classic ARAP~\cite{ARAP_modeling} and the recent APAP~\cite{yoo2024apap}. Since these methods operate on meshes and require handle and anchor points, we adapt our drag instructions accordingly. We map user-specified points to their nearest mesh vertices as handle points. For anchor points, we select vertices in non-editable regions that are farthest from the handles.

\paragraph{Dataset.}

To evaluate the performance of 3D drag-based editing, we construct a comprehensive dataset comprising 100 diverse 3D objects, which are carefully curated from the TRELLIS-500K~\cite{xiang2025structured} and BlenderKit libraries~\cite{blenderkit2024}. For each test case, we provide the base 3D mesh paired with several pre-defined drag instructions. Each instruction explicitly specifies a set of source handle points, their corresponding target locations, and an editable region mask. This dataset provides a standardized and robust testbed for our quantitative and qualitative evaluations.

\begin{figure}[t]
    \centering
\includegraphics[width=0.45\textwidth]{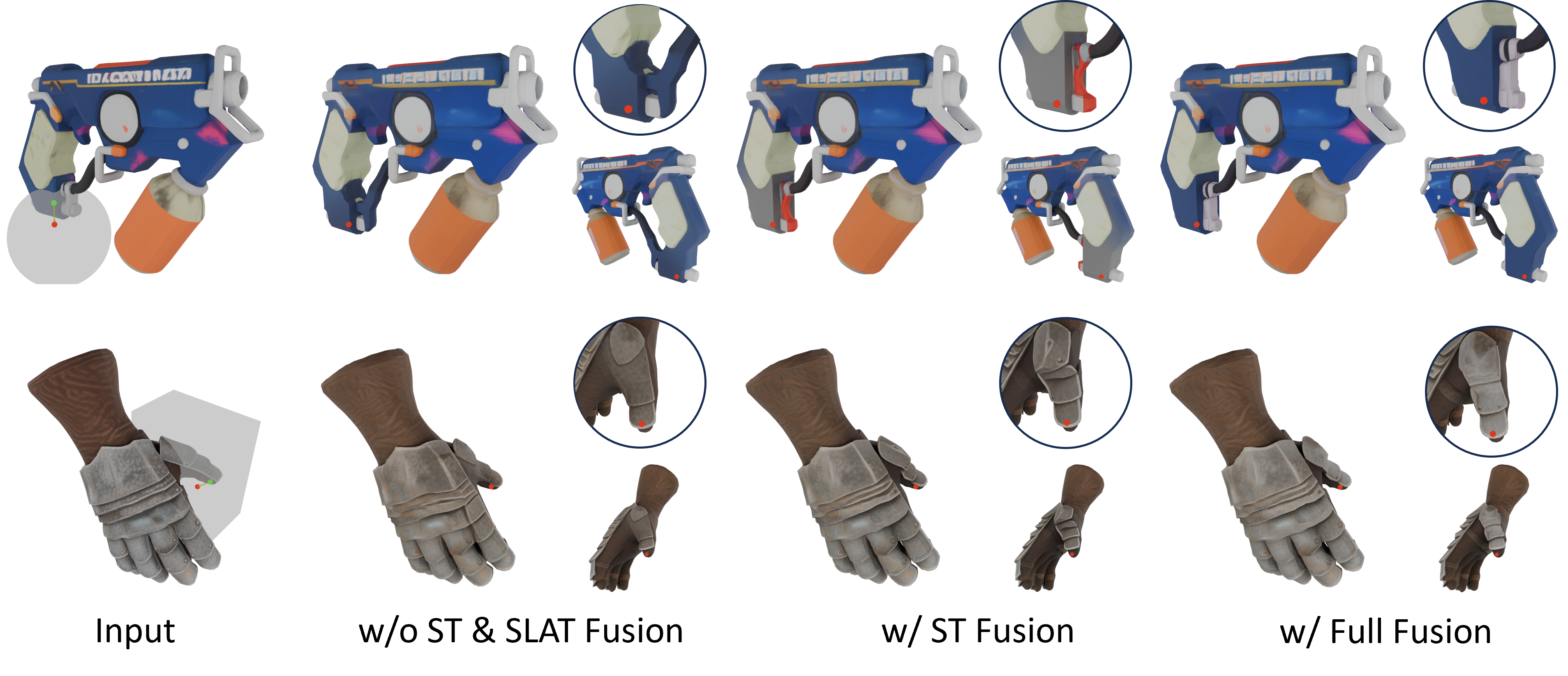}
    \vspace{-4mm}
    \caption{Visualization of logits-level token fusion ablation. Adding ST fusion enables more natural and plausible deformations, while further incorporating SLAT fusion successfully preserves the original texture.}
    \label{fig:ablation_logit}
    \vspace{-5mm}
\end{figure}

\paragraph{Metrics.}

We employ a comprehensive suite of metrics across three key dimensions. (1) \textit{Unedited Region Preservation}: we utilize Chamfer Distance (CD)~\cite{fan2017point} to assess geometric consistency, alongside PSNR, SSIM~\cite{wang2004image}, and LPIPS~\cite{zhang2018unreasonable} to measure texture fidelity. (2) \textit{Drag Accuracy}: To evaluate the editing effectiveness and drag precision, we compute the Dragging Accuracy Index (DAI)~\cite{zhang2025gooddrag} and Mean Distance (MD)~\cite{pan2023drag}. (3) \textit{Overall 3D Quality}: The overall 3D quality and visual coherence of the edited assets are measured using Fréchet Inception Distance (FID)~\cite{heusel2017gans}, Fréchet Video Distance (FVD)~\cite{unterthiner2018towards}, and GPTEval3D~\cite{wu2023gpteval3d}, for which we use GPT-5.2 as the evaluator~\cite{openai2025gpt52}. Additionally, we record and compare the running time of all methods to assess their computational efficiency.

\subsection{Main Results}

\paragraph{Quantitative Evaluation.}

As summarized in Tab.~\ref{tab:comparison}, 3DRAG achieves state-of-the-art performance across all evaluation dimensions. Regarding the preservation of unedited regions, our method achieves the best Chamfer Distance, PSNR, SSIM, and LPIPS, demonstrating strong consistency in both geometry and texture. By leveraging explicit displacement fields and manipulating within a native 3D latent space, 3DRAG achieves outstanding drag precision, as reflected in its optimal DAI and MD scores. The top performance in FID, FVD, and GPTEval3D evaluations confirm that our native 3D generative pipeline attains outstanding visual fidelity and geometric consistency. Furthermore, our method avoids costly per-edit optimization during editing, making it significantly faster than diffusion-based baselines.

\begin{table}[t]
    \centering
    \vspace{-2mm}
    \caption{Quantitative ablation of OA-WTA and logits-level token fusion.}
    \vspace{-4mm}
    \begin{tabular*}{\linewidth}{@{\extracolsep{\fill}}l|cccc} 
    \toprule
    \multirow{2}{*}[-0.75ex]{Method} & \multirow{2}{*}[-0.75ex]{DAI $\downarrow$} & \multirow{2}{*}[-0.75ex]{FVD $\downarrow$} & \multicolumn{2}{c}{User Study $\uparrow$} \\
    \cmidrule(lr){4-5} 
     & & & OA & Fusion \\ 
    \midrule
    w/o OA       & 0.198 & 42.76 & 24.14\% & / \\
    w/o fusion   & 0.224 & 84.09 & / & 10.34\% \\ 
    w/ ST fusion & 0.217 & 43.49 & / & 18.10\% \\ 
    ours         & 0.187 & 42.32 & 75.86\% & 71.55\% \\
    \bottomrule
    \end{tabular*}
    \label{tab:ablation_logit}
\end{table}

\begin{table}[t]
    \centering
    \vspace{-2mm}
    \caption{User Study}
    \vspace{-4mm}
    \resizebox{1.0\linewidth}{!}{
    \begin{tabular}{l|ccc}
    \toprule
     Method & Drag Accuracy $\uparrow$ & Unedited Region Preservation $\uparrow$ & Overall Quality$\uparrow$  \\ 
    \midrule
    3DGS-Drag & 2.03 & 2.17 & 3.00 \\ 
    Drag Your Gaussian & 3.60 & 3.32 & 3.24 \\ 
    ARAP & 1.28 & 1.03 & 1.00 \\
    APAP & 3.17 & 3.54 & 3.01 \\ 
    \midrule
    \textbf{Ours} & \textbf{4.58} & \textbf{4.59} & \textbf{4.40}  \\
    \bottomrule
    \end{tabular}
    }
    \vspace{-2mm}
    \label{tab:user_study}
\end{table}

\paragraph{Qualitative Evaluation.}

As illustrated in Fig.~\ref{fig:compare}, 3DRAG demonstrates exceptional visual superiority in executing drag instructions. The traditional ARAP method is prone to unnatural stretching when applied to meshes with poor topological structures. While the 2D prior–driven APAP alleviates some structural collapse, it often introduces global mesh distortion. Methods that optimize a Gaussian-based representation using SDS often suffer from instability, leading to color shifts as well as blurry artifacts and sparsely scattered Gaussians. In contrast, our method enables coherent and effective drag-based editing while faithfully preserving the object’s original geometric and textural details. Fig.~\ref{fig:gallery} presents additional editing results from 3DRAG.

\subsection{Ablation Study}

\paragraph{Effect of Occupancy-Aware Winner-Takes-All Strategy.}
We ablate the proposed OA-WTA strategy to evaluate its effectiveness in preserving topological integrity. As shown in Fig.~\ref{fig:ablation_geo}, removing the occupancy-aware mechanism leads to severe geometric degradation, evidenced by unnatural tearing and disconnected components in the character’s arms, as well as floating artifacts from the translated chimney. In contrast, by explicitly incorporating a voxel occupancy prior, our strategy prioritizes occupied candidate sources, preventing empty space from eroding valid geometry. This structural improvement is quantitatively validated in Tab.~\ref{tab:ablation_logit}.

\paragraph{Effect of Logits-Level Token Fusion.}
We assess the necessity of our Logits-Level Token Fusion mechanism across the two-stage generative pipeline. As shown in Fig.~\ref{fig:ablation_logit}, disabling fusion in the SLAT stage significantly impairs texture retrieval, leading to noticeable color shifts and hallucinated artifacts, such as the unnatural red patches on the extended gun handle. Further disabling fusion in both stages causes a pronounced degradation in geometric detail, resulting in over-smoothed and ambiguous geometric blends. In contrast, the full method effectively deforms the mesh while faithfully preserving its original identity. These results demonstrate that our design explicitly encourages the model to retrieve and transfer the correct source features. Tab.~\ref{tab:ablation_logit} quantitatively corroborates these consistent improvements across all metrics.

\subsection{User Study}

To assess the perceptual quality of our method, we conducted a user study with 30 participants, 30\% of whom were artists or 3D researchers. We presented the edited 3D assets produced by our method and four baselines in a randomized, blind setup, and asked participants to rank them according to three criteria: Edit Effectiveness (i.e., adherence to the drag instructions), Unedited Region Preservation (i.e., identity consistency within unedited regions), and Overall Object Quality (i.e., visual fidelity and geometric plausibility). The rankings were then aggregated using the Borda count method~\cite{sep-voting-methods}. As reported in Tab.~\ref{tab:user_study}, 3DRAG consistently outperforms all baselines across all three dimensions.

\begin{figure}[t]
    \centering
    \vspace{-2mm}
    \includegraphics[width=0.45\textwidth]{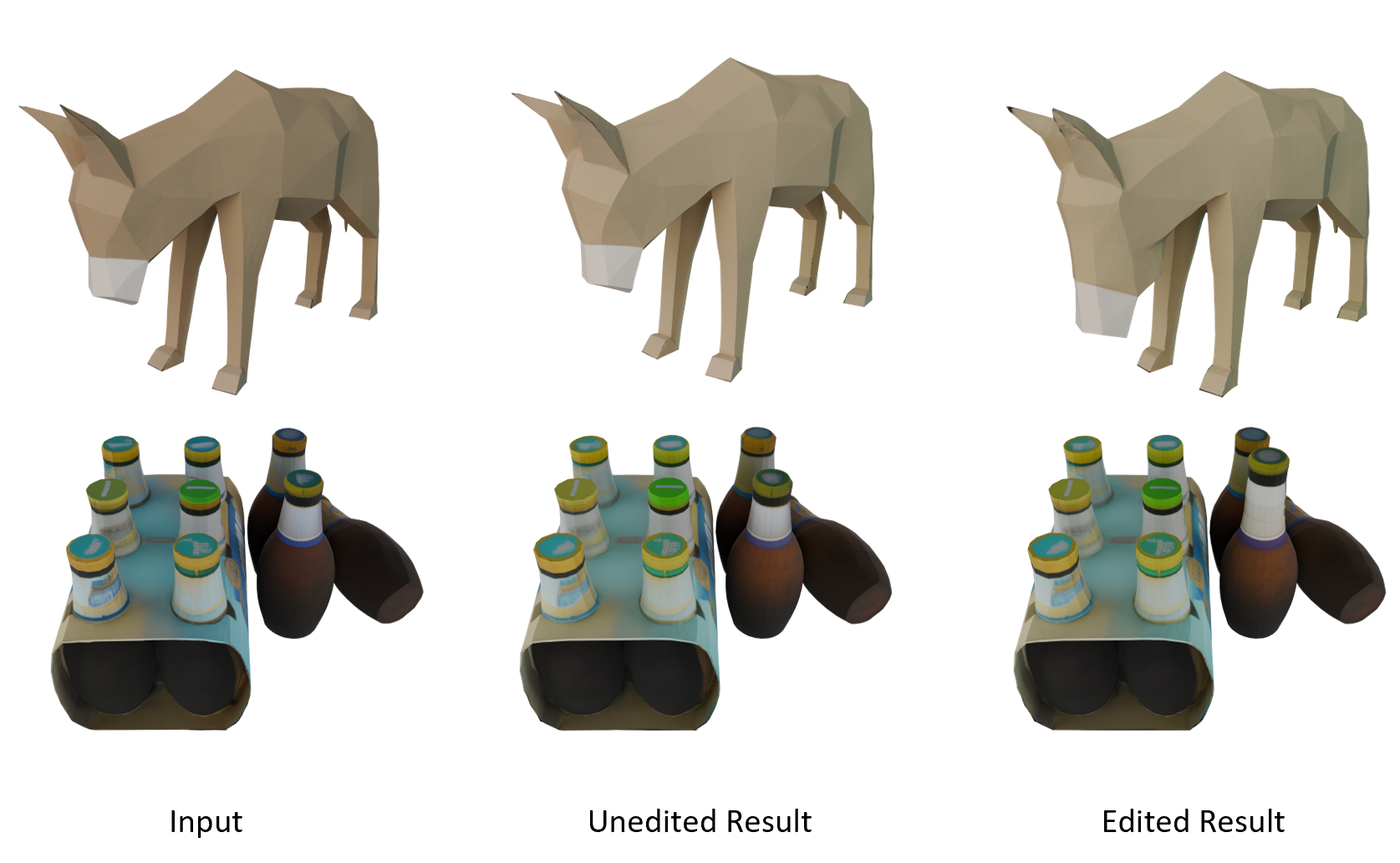}
    \vspace{-4mm}
    \caption{
    Analysis of color shifts. The edited result closely matches the unedited VAE reconstruction, indicating that the slight color shifts are attributable to the TRELLIS VAE.}
    \vspace{-3mm}
    \label{fig:limite}
\end{figure}

\section{Conclusion}

In this paper, we presented 3DRAG, a novel and efficient framework for precise, 3D-native object drag editing. By circumventing costly and unstable 2D diffusion priors, our method executes complex spatial manipulations without costly per-edit optimization. 3DRAG leverages a 3D displacement field equipped with an Occupancy-Aware Winner-Takes-All strategy to robustly preserve topological integrity. Coupled with exact diffusion inversion and our proposed Logits-Level Token Fusion, the framework ensures plausible geometric deformations, seamless texture blending, and the preservation of unedited regions. Experiments demonstrate that 3DRAG achieves state-of-the-art performance in both visual quality and editing accuracy, paving the way for intuitive 3D content creation.

\paragraph{Limitations.}

Our framework has two limitations. First, as analyzed in Fig.~\ref{fig:limite}, slight color shifts may occasionally occur during the editing process. This phenomenon is primarily attributed to the underlying VAE of TRELLIS\cite{xiang2025structured}, which cannot guarantee perfect color consistency upon reconstruction. Quantitatively, the VAE reconstruction yields a mean $\Delta E_{76}$ of 8.14 and a mean CIEDE2000 of 5.59 relative to the original meshes. Second, the spatial precision of the drag operation is inherently bounded by the resolution of the sparse active voxel grid (currently $64^3$). Consequently, applying fine-grained drag instructions to extremely thin structures or highly intricate geometric details may lead to suboptimal deformations. As native 3D foundation models continue to improve in spatial resolution and reconstruction fidelity, we anticipate that our method will naturally benefit from these advancements.

\clearpage

\begin{acks}
This work was supported in part by the National Natural Science Foundation of China under Grant No. W2431046; the Central Guidance on Local Science and Technology Development Fund of China under Grant No. YDZX20253100001001; the Key Laboratory of Intelligent Perception and Human-Machine Collaboration (ShanghaiTech University), Ministry of Education; the Shanghai Frontiers Science Center of Human-Centered Artificial Intelligence; and the HPC Platform of ShanghaiTech University.
\end{acks}

\bibliographystyle{ACM-Reference-Format}
\balance
\bibliography{samples/sample-base}

@String{Computer = "{IEEE} Computer" }

@article{yin2025lazydrag,
  title={LazyDrag: Enabling Stable Drag-Based Editing on Multi-Modal Diffusion Transformers via Explicit Correspondence},
  author={Yin, Zixin and Dai, Xili and Wang, Duomin and Zeng, Xianfang and Ni, Lionel M and Yu, Gang and Shum, Heung-Yeung},
  journal={arXiv preprint arXiv:2509.12203},
  year={2025}
}

@inproceedings{yin2026trainingfree,
  title={Training-Free Text-Guided Color Editing with Multi-Modal Diffusion Transformer},
  author={Yin, Zixin and Dai, Xili and Chen, Ling-Hao and Zhou, Deyu and Wang, Jianan and Wang, Duomin and Yu, Gang and Ni, Lionel M. and Zhang, Lei and Shum, Heung-Yeung},
  booktitle={International Conference on Learning Representations (ICLR)},
  year={2026},
  url={https://openreview.net/forum?id=N1DfzTVuUY}
}

@inproceedings{xiang2025structured,
  title={Structured 3d latents for scalable and versatile 3d generation},
  author={Xiang, Jianfeng and Lv, Zelong and Xu, Sicheng and Deng, Yu and Wang, Ruicheng and Zhang, Bowen and Chen, Dong and Tong, Xin and Yang, Jiaolong},
  booktitle={Proceedings of the IEEE/CVF conference on computer vision and pattern recognition},
  pages={21469--21480},
  year={2025}
}

@article{zhang2024clay,
  title={Clay: A controllable large-scale generative model for creating high-quality 3d assets},
  author={Zhang, Longwen and Wang, Ziyu and Zhang, Qixuan and Qiu, Qiwei and Pang, Anqi and Jiang, Haoran and Yang, Wei and Xu, Lan and Yu, Jingyi},
  journal={ACM Transactions on Graphics (TOG)},
  volume={43},
  number={4},
  pages={1--20},
  year={2024},
  publisher={ACM New York, NY, USA}
}

@inproceedings{livoxhammer,
  title={VoxHammer: Training-Free Precise and Coherent 3D Editing in Native 3D Space},
  author={Li, Lin and Huang, Zehuan and Feng, Haoran and Zhuang, Gengxiong and Chen, Rui and Guo, Chunchao and Sheng, Lu},
  booktitle={Thirteenth International Conference on 3D Vision},
  year={2025}
}

@inproceedings{wang2025taming,
  title={Taming Rectified Flow for Inversion and Editing},
  author={Wang, Jiangshan and Pu, Junfu and Qi, Zhongang and Guo, Jiayi and Ma, Yue and Huang, Nisha and Chen, Yuxin and Li, Xiu and Shan, Ying},
  booktitle={International Conference on Machine Learning},
  pages={64044--64058},
  year={2025},
  organization={PMLR}
}

@article{zhao2024fastdrag,
  title={Fastdrag: Manipulate anything in one step},
  author={Zhao, Xuanjia and Guan, Jian and Fan, Congyi and Xu, Dongli and Lin, Youtian and Pan, Haiwei and Feng, Pengming},
  journal={Advances in Neural Information Processing Systems},
  volume={37},
  pages={74439--74460},
  year={2024}
}

@article{mildenhall2021nerf,
  title={Nerf: Representing scenes as neural radiance fields for view synthesis},
  author={Mildenhall, Ben and Srinivasan, Pratul P and Tancik, Matthew and Barron, Jonathan T and Ramamoorthi, Ravi and Ng, Ren},
  journal={Communications of the ACM},
  volume={65},
  number={1},
  pages={99--106},
  year={2021},
  publisher={ACM New York, NY, USA}
}

@Article{kerbl3Dgaussians,
      author       = {Kerbl, Bernhard and Kopanas, Georgios and Leimk{\"u}hler, Thomas and Drettakis, George},
      title        = {3D Gaussian Splatting for Real-Time Radiance Field Rendering},
      journal      = {ACM Transactions on Graphics},
      number       = {4},
      volume       = {42},
      month        = {July},
      year         = {2023},
      url          = {https://repo-sam.inria.fr/fungraph/3d-gaussian-splatting/}
}

@inproceedings{pooledreamfusion,
  title={DreamFusion: Text-to-3D using 2D Diffusion},
  author={Poole, Ben and Jain, Ajay and Barron, Jonathan T and Mildenhall, Ben},
  booktitle={The Eleventh International Conference on Learning Representations},
  year={2022}
}

@inproceedings{lin2023magic3d,
  title={Magic3d: High-resolution text-to-3d content creation},
  author={Lin, Chen-Hsuan and Gao, Jun and Tang, Luming and Takikawa, Towaki and Zeng, Xiaohui and Huang, Xun and Kreis, Karsten and Fidler, Sanja and Liu, Ming-Yu and Lin, Tsung-Yi},
  booktitle={Proceedings of the IEEE/CVF conference on computer vision and pattern recognition},
  pages={300--309},
  year={2023}
}

@inproceedings{metzer2023latent,
  title={Latent-nerf for shape-guided generation of 3d shapes and textures},
  author={Metzer, Gal and Richardson, Elad and Patashnik, Or and Giryes, Raja and Cohen-Or, Daniel},
  booktitle={Proceedings of the IEEE/CVF conference on computer vision and pattern recognition},
  pages={12663--12673},
  year={2023}
}

@InProceedings{chen2023fantasia3d,
  author={Chen, Rui and Chen, Yongwei and Jiao, Ningxin and Jia, Kui},
  title={Fantasia3D: Disentangling Geometry and Appearance for High-quality Text-to-3D Content Creation},
  booktitle={Proceedings of the IEEE/CVF International Conference on Computer Vision (ICCV)},
  month={October},
  year={2023},
  pages={22246-22256}
}

@inproceedings{wang2023prolificdreamer,
  title={ProlificDreamer: High-Fidelity and Diverse Text-to-3D Generation with Variational Score Distillation},
  author={Zhengyi Wang and Cheng Lu and Yikai Wang and Fan Bao and Chongxuan Li and Hang Su and Jun Zhu},
  booktitle={Advances in Neural Information Processing Systems (NeurIPS)},
  year={2023}
}

@inproceedings{liang2024luciddreamer,
  title={Luciddreamer: Towards high-fidelity text-to-3d generation via interval score matching},
  author={Liang, Yixun and Yang, Xin and Lin, Jiantao and Li, Haodong and Xu, Xiaogang and Chen, Yingcong},
  booktitle={Proceedings of the IEEE/CVF conference on computer vision and pattern recognition},
  pages={6517--6526},
  year={2024}
}

@inproceedings{qiu2024richdreamer,
  title={Richdreamer: A generalizable normal-depth diffusion model for detail richness in text-to-3d},
  author={Qiu, Lingteng and Chen, Guanying and Gu, Xiaodong and Zuo, Qi and Xu, Mutian and Wu, Yushuang and Yuan, Weihao and Dong, Zilong and Bo, Liefeng and Han, Xiaoguang},
  booktitle={Proceedings of the IEEE/CVF conference on computer vision and pattern recognition},
  pages={9914--9925},
  year={2024}
}

@inproceedings{shimvdream,
  title={MVDream: Multi-view Diffusion for 3D Generation},
  author={Shi, Yichun and Wang, Peng and Ye, Jianglong and Mai, Long and Li, Kejie and Yang, Xiao},
  booktitle={The Twelfth International Conference on Learning Representations},
  year={2024}
}

@inproceedings{liu2023zero,
  title={Zero-1-to-3: Zero-shot one image to 3d object},
  author={Liu, Ruoshi and Wu, Rundi and Van Hoorick, Basile and Tokmakov, Pavel and Zakharov, Sergey and Vondrick, Carl},
  booktitle={Proceedings of the IEEE/CVF international conference on computer vision},
  pages={9298--9309},
  year={2023}
}

@inproceedings{long2024wonder3d,
  title={Wonder3d: Single image to 3d using cross-domain diffusion},
  author={Long, Xiaoxiao and Guo, Yuan-Chen and Lin, Cheng and Liu, Yuan and Dou, Zhiyang and Liu, Lingjie and Ma, Yuexin and Zhang, Song-Hai and Habermann, Marc and Theobalt, Christian and others},
  booktitle={Proceedings of the IEEE/CVF conference on computer vision and pattern recognition},
  pages={9970--9980},
  year={2024}
}

@article{wang2023imagedream,
  title={Imagedream: Image-prompt multi-view diffusion for 3d generation},
  author={Wang, Peng and Shi, Yichun},
  journal={arXiv preprint arXiv:2312.02201},
  year={2023}
}

@inproceedings{lu2024direct2,
  title={Direct2.5: Diverse text-to-3d generation via multi-view 2.5d diffusion},
  author={Lu, Yuanxun and Zhang, Jingyang and Li, Shiwei and Fang, Tian and McKinnon, David and Tsin, Yanghai and Quan, Long and Cao, Xun and Yao, Yao},
  booktitle={Proceedings of the IEEE/CVF conference on computer vision and pattern recognition},
  pages={8744--8753},
  year={2024}
}

@inproceedings{gao2022get3d,
title={GET3D: A Generative Model of High Quality 3D Textured Shapes Learned from Images},
author={Jun Gao and Tianchang Shen and Zian Wang and Wenzheng Chen and Kangxue Yin
and Daiqing Li and Or Litany and Zan Gojcic and Sanja Fidler},
booktitle={Advances In Neural Information Processing Systems},
year={2022}
}

@inproceedings{honglrm,
  title={LRM: Large Reconstruction Model for Single Image to 3D},
  author={Hong, Yicong and Zhang, Kai and Gu, Jiuxiang and Bi, Sai and Zhou, Yang and Liu, Difan and Liu, Feng and Sunkavalli, Kalyan and Bui, Trung and Tan, Hao},
  booktitle={The Twelfth International Conference on Learning Representations},
  year={2024}
}

@inproceedings{zhang2024gs,
  title={GS-LRM: Large Reconstruction Model for 3D Gaussian Splatting},
  author={Zhang, Kai and Bi, Sai and Tan, Hao and Xiangli, Yuanbo and Zhao, Nanxuan and Sunkavalli, Kalyan and Xu, Zexiang},
  booktitle={European Conference on Computer Vision},
  pages={1--19},
  year={2024}
}

@article{xu2024instantmesh,
  title={InstantMesh: Efficient 3D Mesh Generation from a Single Image with Sparse-view Large Reconstruction Models},
  author={Xu, Jiale and Cheng, Weihao and Gao, Yiming and Wang, Xintao and Gao, Shenghua and Shan, Ying},
  journal={arXiv preprint arXiv:2404.07191},
  year={2024}
}

@article{TripoSR2024,
  title={TripoSR: Fast 3D Object Reconstruction from a Single Image},
  author={Tochilkin, Dmitry and Pankratz, David and Liu, Zexiang and Huang, Zixuan and Letts, Adam and Li, Yangguang and Liang, Ding and Laforte, Christian and Jampani, Varun and Cao, Yan-Pei},
  journal={arXiv preprint arXiv:2403.02151},
  year={2024}
}

@article{li2025triposg,
  title={Triposg: High-fidelity 3d shape synthesis using large-scale rectified flow models},
  author={Li, Yangguang and Zou, Zi-Xin and Liu, Zexiang and Wang, Dehu and Liang, Yuan and Yu, Zhipeng and Liu, Xingchao and Guo, Yuan-Chen and Liang, Ding and Ouyang, Wanli and others},
  journal={IEEE Transactions on Pattern Analysis and Machine Intelligence},
  year={2025},
  publisher={IEEE}
}

@inproceedings{pan2023drag,
  title={Drag your gan: Interactive point-based manipulation on the generative image manifold},
  author={Pan, Xingang and Tewari, Ayush and Leimk{\"u}hler, Thomas and Liu, Lingjie and Meka, Abhimitra and Theobalt, Christian},
  booktitle={ACM SIGGRAPH 2023 conference proceedings},
  pages={1--11},
  year={2023}
}

@inproceedings{shi2024dragdiffusion,
  title={Dragdiffusion: Harnessing diffusion models for interactive point-based image editing},
  author={Shi, Yujun and Xue, Chuhui and Liew, Jun Hao and Pan, Jiachun and Yan, Hanshu and Zhang, Wenqing and Tan, Vincent YF and Bai, Song},
  booktitle={Proceedings of the IEEE/CVF conference on computer vision and pattern recognition},
  pages={8839--8849},
  year={2024}
}

@inproceedings{moudragondiffusion,
  title={DragonDiffusion: Enabling Drag-style Manipulation on Diffusion Models},
  author={Mou, Chong and Wang, Xintao and Song, Jiechong and Shan, Ying and Zhang, Jian},
  booktitle={The Twelfth International Conference on Learning Representations},
  year={2024}
}

@inproceedings{mou2024diffeditor,
  title={Diffeditor: Boosting accuracy and flexibility on diffusion-based image editing},
  author={Mou, Chong and Wang, Xintao and Song, Jiechong and Shan, Ying and Zhang, Jian},
  booktitle={Proceedings of the IEEE/CVF Conference on Computer Vision and Pattern Recognition},
  pages={8488--8497},
  year={2024}
}

@inproceedings{ling2024freedrag,
  title={Freedrag: Feature dragging for reliable point-based image editing},
  author={Ling, Pengyang and Chen, Lin and Zhang, Pan and Chen, Huaian and Jin, Yi and Zheng, Jinjin},
  booktitle={Proceedings of the IEEE/CVF conference on computer vision and pattern recognition},
  pages={6860--6870},
  year={2024}
}

@inproceedings{cui2024stabledrag,
  title={StableDrag: Stable Dragging for Point-Based Image Editing},
  author={Cui, Yutao and Zhao, Xiaotong and Zhang, Guozhen and Cao, Shengming and Ma, Kai and Wang, Limin},
  booktitle={European Conference on Computer Vision},
  pages={340--356},
  year={2024}
}

@inproceedings{shin2024instantdrag,
      title     = {{InstantDrag: Improving Interactivity in Drag-based Image Editing}},
      author    = {Shin, Joonghyuk and Choi, Daehyeon and Park, Jaesik},
      booktitle = {ACM SIGGRAPH Asia 2024 Conference Proceedings},
      year      = {2024},
      pages     = {1--10},
}

@inproceedings{shi2025lightningdrag,
  title={LightningDrag: Lightning Fast and Accurate Drag-based Image Editing Emerging from Videos},
  author={Shi, Yujun and Liew, Jun Hao and Yan, Hanshu and Tan, Vincent and Feng, Jiashi},
  booktitle={International Conference on Machine Learning},
  pages={54981--54993},
  year={2025},
  organization={PMLR}
}

@inproceedings{koo2025flowdrag,
  title={FlowDrag: 3D-aware Drag-based Image Editing with Mesh-guided Deformation Vector Flow Fields},
  author={Koo, Gwanhyeong and Yoon, Sunjae and Lee, Younghwan and Hong, Ji Woo and Yoo, Chang D},
  booktitle={International Conference on Machine Learning},
  pages={31467--31485},
  year={2025},
  organization={PMLR}
}

@inproceedings{liu2024drag,
  title={Drag your noise: Interactive point-based editing via diffusion semantic propagation},
  author={Liu, Haofeng and Xu, Chenshu and Yang, Yifei and Zeng, Lihua and He, Shengfeng},
  booktitle={Proceedings of the IEEE/CVF conference on computer vision and pattern recognition},
  pages={6743--6752},
  year={2024}
}

@inproceedings{liu2021editing,
  title={Editing conditional radiance fields},
  author={Liu, Steven and Zhang, Xiuming and Zhang, Zhoutong and Zhang, Richard and Zhu, Jun-Yan and Russell, Bryan},
  booktitle={Proceedings of the IEEE/CVF international conference on computer vision},
  pages={5773--5783},
  year={2021}
}

@inproceedings{mirzaei2023spin,
  title={Spin-nerf: Multiview segmentation and perceptual inpainting with neural radiance fields},
  author={Mirzaei, Ashkan and Aumentado-Armstrong, Tristan and Derpanis, Konstantinos G and Kelly, Jonathan and Brubaker, Marcus A and Gilitschenski, Igor and Levinshtein, Alex},
  booktitle={Proceedings of the IEEE/CVF Conference on Computer Vision and Pattern Recognition},
  pages={20669--20679},
  year={2023}
}

@misc{
  yin2023ornerf,
  title={OR-NeRF: Object Removing from 3D Scenes Guided by Multiview Segmentation with Neural Radiance Fields}, 
  author={Youtan Yin and Zhoujie Fu and Fan Yang and Guosheng Lin},
  year={2023},
  eprint={2305.10503},
  archivePrefix={arXiv},
  primaryClass={cs.CV}
}

@article{zhuang2024tip,
  title={Tip-editor: An accurate 3d editor following both text-prompts and image-prompts},
  author={Zhuang, Jingyu and Kang, Di and Cao, Yan-Pei and Li, Guanbin and Lin, Liang and Shan, Ying},
  journal={ACM Transactions on Graphics (ToG)},
  volume={43},
  number={4},
  pages={1--12},
  year={2024},
  publisher={ACM New York, NY, USA}
}

@inproceedings{park2024ed,
  title={ED-NERF: EFFICIENT TEXT-GUIDED EDITING OF 3D SCENE WITH LATENT SPACE NERF},
  author={Park, Jangho and Kwon, Gihyun and Ye, Jong Chul},
  booktitle={12th International Conference on Learning Representations, ICLR 2024},
  year={2024}
}

@inproceedings{chen2023neuraleditor,
  title={Neuraleditor: Editing neural radiance fields via manipulating point clouds},
  author={Chen, Jun-Kun and Lyu, Jipeng and Wang, Yu-Xiong},
  booktitle={Proceedings of the IEEE/CVF conference on computer vision and pattern recognition},
  pages={12439--12448},
  year={2023}
}

@inproceedings{haque2023instruct,
  title={Instruct-nerf2nerf: Editing 3d scenes with instructions},
  author={Haque, Ayaan and Tancik, Matthew and Efros, Alexei A and Holynski, Aleksander and Kanazawa, Angjoo},
  booktitle={Proceedings of the IEEE/CVF international conference on computer vision},
  pages={19740--19750},
  year={2023}
}

@inproceedings{chen2024gaussianeditor,
  title={Gaussianeditor: Swift and controllable 3d editing with gaussian splatting},
  author={Chen, Yiwen and Chen, Zilong and Zhang, Chi and Wang, Feng and Yang, Xiaofeng and Wang, Yikai and Cai, Zhongang and Yang, Lei and Liu, Huaping and Lin, Guosheng},
  booktitle={Proceedings of the IEEE/CVF conference on computer vision and pattern recognition},
  pages={21476--21485},
  year={2024}
}

@inproceedings{wu2024gaussctrl,
  title={GaussCtrl: Multi-view Consistent Text-Driven 3D Gaussian Splatting Editing},
  author={Wu, Jing and Bian, Jia-Wang and Li, Xinghui and Wang, Guangrun and Reid, Ian and Torr, Philip and Prisacariu, Victor Adrian},
  booktitle={European Conference on Computer Vision},
  pages={55--71},
  year={2024}
}

@inproceedings{chen2024dge,
  title={DGE: Direct Gaussian 3D Editing by Consistent Multi-view Editing},
  author={Chen, Minghao and Laina, Iro and Vedaldi, Andrea},
  booktitle={European Conference on Computer Vision},
  pages={74--92},
  year={2024}
}

@inproceedings{wen2025intergsedit,
  title={Intergsedit: Interactive 3d gaussian splatting editing with 3d geometry-consistent attention prior},
  author={Wen, Minghao and Wu, Shengjie and Wang, Kangkan and Liang, Dong},
  booktitle={Proceedings of the IEEE/CVF International Conference on Computer Vision},
  pages={26136--26145},
  year={2025}
}

@inproceedings{sun2024gseditpro,
  title={GSEditPro: 3D Gaussian Splatting Editing with Attention-based Progressive Localization},
  author={Sun, Yanhao and Tian, RunZe and Han, Xiao and Liu, XinYao and Zhang, Yan and Xu, Kai},
  booktitle={Computer Graphics Forum},
  volume={43},
  number={7},
  year={2024}
}

@inproceedings{qu2025drag,
  title={Drag your gaussian: Effective drag-based editing with score distillation for 3d gaussian splatting},
  author={Qu, Yansong and Chen, Dian and Li, Xinyang and Li, Xiaofan and Zhang, Shengchuan and Cao, Liujuan and Ji, Rongrong},
  booktitle={Proceedings of the Special Interest Group on Computer Graphics and Interactive Techniques Conference Conference Papers},
  pages={1--12},
  year={2025}
}

@inproceedings{dong3dgs,
  title={3DGS-Drag: Dragging Gaussians for Intuitive Point-Based 3D Editing},
  author={Dong, Jiahua and Wang, Yu-Xiong},
  booktitle={The Thirteenth International Conference on Learning Representations},
  year={2025}
}

@inproceedings{li2025cmd,
  title={Cmd: Controllable multiview diffusion for 3d editing and progressive generation},
  author={Li, Peng and Ma, Suizhi and Chen, Jialiang and Liu, Yuan and Zhang, Congyi and Xue, Wei and Luo, Wenhan and Sheffer, Alla and Wang, Wenping and Guo, Yike},
  booktitle={Proceedings of the Special Interest Group on Computer Graphics and Interactive Techniques Conference Conference Papers},
  pages={1--10},
  year={2025}
}

@article{chi2026vinedresser3d,
  title={Vinedresser3D: Agentic Text-guided 3D Editing},
  author={Chi, Yankuan and Li, Xiang and Huang, Zixuan and Rehg, James M},
  journal={arXiv preprint arXiv:2602.19542},
  year={2026}
}

@inproceedings{ARAP_modeling,
author = {Olga Sorkine and Marc Alexa},
title = {As-Rigid-As-Possible Surface Modeling},
booktitle = {Proceedings of EUROGRAPHICS/ACM SIGGRAPH Symposium on Geometry Processing},
year = {2007},
pages = {109--116},
}

@inproceedings{yoo2024apap,
  title = {{As-Plausible-As-Possible: Plausibility-Aware Mesh Deformation Using 2D Diffusion Priors}},
  author = {Yoo, Seungwoo and Kim, Kunho and Kim, Vladimir G. and Sung, Minhyuk},
  booktitle = {CVPR},
  year = {2024},
}

@inproceedings{fan2017point,
  title={A point set generation network for 3d object reconstruction from a single image},
  author={Fan, Haoqiang and Su, Hao and Guibas, Leonidas J},
  booktitle={Proceedings of the IEEE conference on computer vision and pattern recognition},
  pages={605--613},
  year={2017}
}

@article{wang2004image,
  title={Image quality assessment: from error visibility to structural similarity},
  author={Wang, Zhou and Bovik, Alan C and Sheikh, Hamid R and Simoncelli, Eero P},
  journal={IEEE transactions on image processing},
  volume={13},
  number={4},
  pages={600--612},
  year={2004},
  publisher={IEEE}
}

@inproceedings{zhang2018unreasonable,
  title={The unreasonable effectiveness of deep features as a perceptual metric},
  author={Zhang, Richard and Isola, Phillip and Efros, Alexei A and Shechtman, Eli and Wang, Oliver},
  booktitle={Proceedings of the IEEE conference on computer vision and pattern recognition},
  pages={586--595},
  year={2018}
}

@inproceedings{zhang2025gooddrag,
  title={{GoodDrag}: Towards Good Practices for Drag Editing with Diffusion Models},
  author={Zhang, Zewei and Liu, Huan and Chen, Jun and Xu, Xiangyu},
  booktitle={The Thirteenth International Conference on Learning Representations},
  year={2025}
}

@article{heusel2017gans,
  title={Gans trained by a two time-scale update rule converge to a local nash equilibrium},
  author={Heusel, Martin and Ramsauer, Hubert and Unterthiner, Thomas and Nessler, Bernhard and Hochreiter, Sepp},
  journal={Advances in neural information processing systems},
  volume={30},
  year={2017}
}

@article{unterthiner2018towards,
  title={Towards accurate generative models of video: A new metric \& challenges},
  author={Unterthiner, Thomas and Van Steenkiste, Sjoerd and Kurach, Karol and Marinier, Raphael and Michalski, Marcin and Gelly, Sylvain},
  journal={arXiv preprint arXiv:1812.01717},
  year={2018}
}

@inproceedings{wu2023gpteval3d,
   author = {Tong Wu and Guandao Yang and Zhibing Li and Kai Zhang and
             Ziwei Liu and Leonidas Guibas and Dahua Lin and Gordon Wetzstein},
   title = {GPT-4V(ision) is a Human-Aligned Evaluator for Text-to-3D Generation},
   booktitle = {CVPR},
   year = {2024},
}

@misc{blenderkit2024,
  author       = {{BlenderKit}},
  title        = {BlenderKit: Download free 3D models, textures and Blender assets},
  year         = {2026},
  howpublished = {\url{https://www.blenderkit.com}},
  note         = {Accessed: 2026-03-25}
}

@InCollection{sep-voting-methods,
	author       =	{Pacuit, Eric},
	title        =	{{Voting Methods}},
	booktitle    =	{The {Stanford} Encyclopedia of Philosophy},
	editor       =	{Edward N. Zalta and Uri Nodelman},
	howpublished =	{\url{https://plato.stanford.edu/archives/spr2026/entries/voting-methods/}},
	year         =	{2026},
	edition      =	{{S}pring 2026},
	publisher    =	{Metaphysics Research Lab, Stanford University}
}

@inproceedings{cao2023masactrl,
  title={Masactrl: Tuning-free mutual self-attention control for consistent image synthesis and editing},
  author={Cao, Mingdeng and Wang, Xintao and Qi, Zhongang and Shan, Ying and Qie, Xiaohu and Zheng, Yinqiang},
  booktitle={Proceedings of the IEEE/CVF international conference on computer vision},
  pages={22560--22570},
  year={2023}
}

@inproceedings{peebles2023scalable,
  title={Scalable diffusion models with transformers},
  author={Peebles, William and Xie, Saining},
  booktitle={Proceedings of the IEEE/CVF international conference on computer vision},
  pages={4195--4205},
  year={2023}
}

@inproceedings{salimans2022progressive,
title={Progressive Distillation for Fast Sampling of Diffusion Models},
author={Tim Salimans and Jonathan Ho},
booktitle={International Conference on Learning Representations},
year={2022},
url={https://openreview.net/forum?id=TIdIXIpzhoI}
}

@inproceedings{ruiz2023dreambooth,
  title={Dreambooth: Fine tuning text-to-image diffusion models for subject-driven generation},
  author={Ruiz, Nataniel and Li, Yuanzhen and Jampani, Varun and Pritch, Yael and Rubinstein, Michael and Aberman, Kfir},
  booktitle={Proceedings of the IEEE/CVF conference on computer vision and pattern recognition},
  pages={22500--22510},
  year={2023}
}

@inproceedings{hong2023debiasing,
title={Debiasing Scores and Prompts of 2D Diffusion for View-consistent Text-to-3D Generation},
author={Susung Hong and Donghoon Ahn and Seungryong Kim},
booktitle={Thirty-seventh Conference on Neural Information Processing Systems},
year={2023},
url={https://openreview.net/forum?id=jgIrJeHHlz}
}

@article{su2024roformer,
  title={RoFormer: Enhanced transformer with Rotary Position Embedding},
  author={Su, Jianlin and Ahmed, Murtadha and Lu, Yu and Pan, Shengfeng and Bo, Wen and Liu, Yunfeng},
  journal={Neurocomputing},
  volume={568},
  pages={127063},
  year={2024},
  publisher={Elsevier}
}

@inproceedings{ho2021classifierfree,
title={Classifier-Free Diffusion Guidance},
author={Jonathan Ho and Tim Salimans},
booktitle={NeurIPS 2021 Workshop on Deep Generative Models and Downstream Applications},
year={2021},
url={https://openreview.net/forum?id=qw8AKxfYbI}
}

@misc{openai2025gpt52,
  author       = {{OpenAI}},
  title        = {Introducing GPT-5.2},
  year         = {2025},
  month        = dec,
  howpublished = {\url{https://openai.com/index/introducing-gpt-5-2/}},
  note         = {Accessed: 2026-08-01}
}

\end{document}